\documentclass[aps,twocolumn,showpacs,superscriptaddress]{revtex4-2}
\usepackage{amsmath,amssymb}
\usepackage{graphics,graphicx}
\usepackage{dcolumn,bm}
\usepackage{breqn}
\usepackage{psfrag}
\usepackage{color}
\usepackage{enumitem}
\usepackage [english]{babel}
\usepackage [autostyle, english = american]{csquotes}
\MakeOuterQuote{"}
\usepackage{blkarray}
\newcommand{\cu}
{\affiliation{Department of Physics, University of Calcutta,
92 Acharya Prafulla Chandra Road, Kolkata 700009, India.}}

\begin{document}

\title{ Random walk with multiple memory channels: a new paradigm }

\author{Surajit Saha}%
\cu


\normalsize  

\begin{abstract}
A new class of one-dimensional, discrete time random walk model with memory, termed "Random walk with $n$ memory channels" (RW$n$MC) is proposed. In this model the information of $n$  ($n\in \mathbb{Z}$)  previous steps from the walker's entire history are needed to decide future step. Exact calculation of the mean and variance of position of the RW2MC ($n=2$) has been done which shows that it can lead to asymptotic diffusive and superdiffusive behavior in different parameter regimes. A connection between RW$n$MC and  Pólya type urn model evolving by $n$ drawings has also been reported. This connection for the RW2MC is discussed in detail which suggests the applicability of RW$2$MC in many population dynamics model with multiple competing species.

\end{abstract}


\maketitle


The term “long range memory” which is sometimes called “long term persistence”—implies that there is non-negligible dependence between the present and the points in the past in a dynamical process. Long range memory plays
a significant role in many fields like astrophysics \cite{sun1,solar1}, atmospheric science \cite{atm}, genomics \cite{gen}, financial markets \cite{eco1,eco2}, complex networks \cite{com1,com2}, geophysics \citep{geo1,geo2} etc. 
  \\
In this communication we present a  non-markovian  discrete time random walk model where the random increment at time step $t$ depends on the complete history of the process. We term this walk as "Random walk with $n$ memory channels" (RW$n$MC) where $n$ is a integer which indicates the number of memory channels. At time $t$, the walker can choose $n$ ($n \geq 1$) previous steps from its entire history with equal a priori probabilities, based on which it makes the decision in the next step by following certain rule. \\
  RW$2$MC presented here offers the great advantage of analytical tractability. Exact calculation of first and second moment of position shows that it exhibits asymptotic diffusive and superdiffusive  behavior in different parameter regimes. Within superdiffusive regime a nontrivial ballistic behavior can be seen for a certain parameter range which has not been noticed earlier for any random walk with complete memory of its history.\\ 
  
  We have also shown that RW$2$MC have identical distribution with a proposed Pólya type urn model, discussed in detail later in the paper. In brief the evolution of a composition of the urn can be transformed into statements about the evolution of position of the RW$2$MC. We then generalise this connection to urn model for $n>2$. The urn model equivalent to RW$2$MC has not been considered earlier to the best of the author's knowledge. \\
   
  
\textit{Construction of RW$n$MC:}
Let us define a RW$2$MC in 1D lattice.
At each step the walker has three options:
it can move to the nearest-neighbor site to its right, to the site on its left, or it can remain at its present location. Denoting the position of the walker at time  $t$ as $X_t$ , one can write a probabilistic
recurrence relation:
\begin{equation}
X_{t+1}=X_t+\sigma_{t+1},
\label{eq1}
\end{equation}
where $\sigma_{t+1}$ is a random number which can take one of the
values $-1$, $0$, or $+1$ according to the following rule: 
 \begin{equation}
\sigma_{t+1}={\mu}^{(1)}_{t+1}\lambda^{(1)}_{t+1}+ \mu^{(2)}_{t+1}\lambda^{(2)}_{t+1}.
\label{eq2}
\end{equation}
Here $\lambda^{(1)}_{t+1}$ and $\lambda^{(2)}_{t+1}$ are  random numbers which  depend on the entire history of the walk $\{\sigma_t \}=(\sigma_1,...,\sigma_t)$ as follows: two random previous times $k_1$  and $k_2$ ($ k_1 \neq k_2$) between $1$ and $t$ are chosen with uniform probability such that $\lambda^{(1)}_{t+1}=\sigma_{k_1}$ and  $\lambda^{(2)}_{t+1}=\sigma_{k_2}$. 
 
${\mu}^{(1)}_{t+1}$ and ${\mu}^{(2)}_{t+1}$ in Eq.(\ref{eq2}) are two random variables which can take two possible
values, $+1$ or $-1$. We define a parameter $p$ as the probability of occurrence of positive
value of $\mu^{(1)}_{t+1}$ and ${\mu}^{(2)}_{t+1}$. If by following the above equation, the value of $\sigma_{t+1}$
 becomes higher (lower) than +1 (-1) then it is made equal to +1 (-1).
 The process is initiated as follows: at  $t = 1$ and $2$, we allow the walker to take steps to the right with
probability $s$ and to the left with probability $1 - s$. So, the first two steps excludes the possibility that the walker may not move. Hence at $t=2$, possible positions are $\pm 2$ and $0$. If the walker starts from initial position $X_0 = 0$, then the position of the walker at time $t>0$ is given by
\begin{equation}
X_t=\sum^t_{k=1}\sigma_k.
\end{equation}
 \\ 
 One can construct a RW$n$MC ($n>2$) model by increasing the terms in R.H.S of Eq.(\ref{eq2}), i.e the walker can choose  uniformly at random  $n$ ($n > 2$) steps from its entire history $\lbrace \sigma_1,\sigma_2,...,\sigma_t \rbrace$ for  making $\sigma_{t+1}$ such that
\begin{equation}
\sigma_{t+1}=\sum^{n}_{i=1} \mu^{(i)}_{t+1}\lambda^{(i)}_{t+1},
\label{rwnmc}
\end{equation}
where $\sigma_{t+1}$ is again taken to be within $\pm 1$. \\



\begin{figure}
\hspace*{-3cm}    
\includegraphics[scale=0.45]{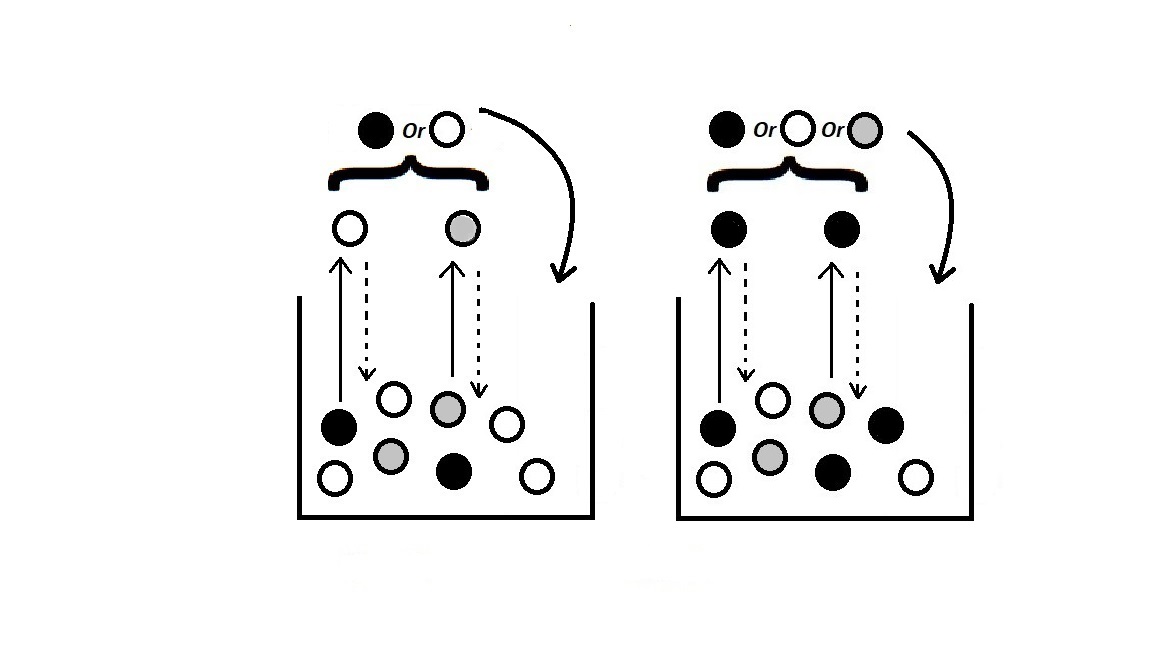}
\caption{Pólya type urn containing balls of three different colors and evolving by two drawings according to the mean replacement matrix (\ref{matrix}).}
\label{u2}
\end{figure}

\textit{Equivalent urn model:} Let us now introduce a discrete-time urn model with balls of three colors. Assume the three colors to be black (B), white (W) and gray (G). Suppose one randomly draws two balls at a time (Fig.\ref{u2}). The possible outcome of any draw are $[B,B]$, $[B,W] $, $[W,W]$, $[G,G]$, $[G,W]$ and $[G,B]$. After observing, the drawn pair is replaced into the urn and a B / W / G ball added into it. Which ball will be added into the urn depends on the drawn pair according to the following mean replacement matrix:
 \vspace*{-0.1cm}
\begin{equation}
\begin{blockarray}{cccc}
  &\:\: B \:\:\:\:\:\:\:& W & G \\
\begin{block}{c(ccc)}
  BB\:\: \,  & a^2\:\:& b^2\:\: & 2ab\:  \\
  BW\:\:\:  & ab & ab & a^2 +b^2\\
  BG \:\:\,  & a & b & 0  \\
  WW \:\:\,  & b^2 & a^2 & 2ab  \\
  WG \:\:\,  & b & a & 0  \\
  GG \:\:\,  & 0 & 0 & 1  \\
 \end{block}
\end{blockarray}
\label{matrix}
\end{equation}
 \vspace*{-0.50cm}
\\
Here in general $a+b=1$. The elements of the 
matrix represent the conditional probability of the colored  ball added in each step. The urn described above belongs to the Pólya type urns evolving by two drawing with randomized replacement rules \cite{mahmoud}. If we associate $+1$ to black balls, $-1$ to white balls and $0$ to gray balls then  the mean replacement matrix (\ref{matrix}) can be obtained for the choice $a=p$ and $b=1-p$ from Eq.(\ref{eq2}).\\
The composition of the urn at time $t$ is
given by a set  $N_t = (N^B_t ,N^W_t,N^G_t)$ where $N^B_t$, $N^W_t$ and $N^G_t$ counts the number of black, white and grey balls respectively.
We restrict ourselves to selected starting compositions with balls of 2 colors only such that at time $t=0$,  $N_0$ can take values either $(1,1,0)$ or $(2,0,0)$ or $(0,2,0)$. This restriction is necessary to establish connection with the RW$2$MC discussed earlier.
\\ 
 One can show that for a particular $p$, if $X_t$ is the position of the walker started from $X_0=0$ such that $X_2=N^B_0 - N^W_0$, then
\begin{equation}
X_{t+2}=_{d} N^{B}_t-N^{W}_t,
\end{equation}
where $=_{d}$ implies equality in law or equality in distribution. That is to say,  the difference between the number of black and white balls in the urn at time $t$ follows the same distribution like position of the RW2MC at time $t+2$ with position at $t=2$ equaling $N^B_0 - N^W_0$.
 \\
Now generalising the above process one can verify that the equivalent urn model of RWnMC ($n>2$) will be a Polya type urn model containing balls of three colors and evolving by $n$ drawings with suitable initial conditions. One can obtain the corresponding mean replacement matrix from Eq.(\ref{rwnmc}) by following same procedure.





 \textit{Calculation of moments}: The mean displacement and the
mean-square displacement as functions of time characterizes the nature of the
motion of the walker. For RW2MC both of
it can be calculated analytically. But here we include the possibility of $k_1=k_2$ i.e. walker can randomly select same two steps from its history. Since the probability of occurrence of such event is very low ($\frac{1}{t}$), inclusion of such possibility does not cause any notable difference in the assymptotic limit ($t\gg1$) from the case $k_1\neq k_2$.\\
Let us introduce shorthand notations $\gamma_{t+1}$ and $\gamma^{\prime}_{t+1}$,
 where $\gamma_{t+1}=\mu^{(1)}_{t+1}\lambda^{(1)}_{t+1}$ and $\gamma^{\prime}_{t+1}=\mu^{(2)}_{t+1}\lambda^{(2)}_{t+1}$.\\
 
 Suppose $f_1(p,t)$, $f_{-1}(p,t)$ and $f_0(p,t)$ are the fraction of $+1$, $-1$ and $0$ steps respectivly in $\lbrace\sigma_t\rbrace$. Then one must have
\begin{equation}
f_1(p,t)+f_{-1}(p,t)+f_0(p,t)=1.
\label{pc}
 \end{equation}
For a given history $\lbrace\sigma_t\rbrace$, the conditional probability that $\gamma_{t+1} =\pm 1,0 $, for $t > 2$, can be written as
\begin{eqnarray*}
&&P(\gamma_{t+1}=\pm 1)=\frac{1}{2t}\sum_{k=1}^t(1-(1-\sigma_k^2)+(2p-1)\sigma_k \gamma_{t+1}),\\
&&P(\gamma_{t+1}=0)=\frac{1}{2t}\sum_{k=1}^t 2(1-\sigma_k^2).
\end{eqnarray*}
Similar equations can be written for $P(\gamma^{\prime}_{t+1}=\pm 1) $ and $P(\gamma^{\prime}_{t+1}=0)$. Now for a given history $\lbrace\sigma_t\rbrace$ and $\gamma^{\prime}_{t+1}$ the conditional probabilities that the increment $\sigma_{t+1}$ takes the value $\pm 1$ and $0$ are 
\vspace*{-0.5 cm}
\begin{multline}
  P(\sigma_{t+1}= 1\vert \lbrace\sigma_t\rbrace,\gamma^{\prime}_{t+1})=(1-\frac{\gamma^{\prime}_{t+1}}{2})(1+\gamma^{\prime}_{t+1})\\
  \times[P(\gamma_{t+1}=+1)+
 \gamma^{\prime}_{t+1}P(\gamma_{t+1}=0)],
 \end{multline}
 \vspace*{-0.8 cm}
\begin{multline}
P(\sigma_{t+1}=-1\vert \lbrace\sigma_t\rbrace,\gamma^{\prime}_{t+1})=(1+\frac{\gamma^{\prime}_{t+1}}{2})(1-\gamma^{\prime}_{t+1})\\
\times[P(\gamma_{t+1}=-1)-\gamma^{\prime}_{t+1} P(\gamma_{t+1}=0)],
\end{multline}
\vspace*{-0.2cm}
If $\gamma^{\prime}_{t+1}=\pm 1$ then,
\begin{equation}
P(\sigma_{t+1}=0\vert \lbrace\sigma_t\rbrace,\gamma^{\prime}_{t+1})=\gamma^{\prime}_{t+1} P(\gamma_{t+1}=-\gamma^{\prime}_{t+1}), 
\end{equation}
and if $\gamma^{\prime}_{t+1}=0$ then,
\begin{equation}
P(\sigma_{t+1}=0\vert \lbrace\sigma_t\rbrace,\gamma^{\prime}_{t+1})=P(\gamma_{t+1}=0). 
\end{equation}
These follows from the dynamics (\ref{eq2}) of the process. So,
The conditional mean increment is given by
\begin{equation}
\langle\sigma_{t+1}\vert \lbrace\sigma_t\rbrace  ,\gamma^{\prime}_{t+1}\rangle=P(\sigma_{t+1}= 1\vert \lbrace\sigma_t\rbrace)-P(\sigma_{t+1}=-1\vert \lbrace\sigma_t\rbrace).
\end{equation}
 Averaging over $\gamma^{\prime}_{t+1}$ one can obtain
\begin{equation}
\langle\sigma_{t+1}\vert \lbrace\sigma_t\rbrace\rangle=\frac{(2p-1)(1+f_0(p,t))}{t}X_t,
\end{equation}
which, on averaging over all the histories, gives the following expression for mean value
\begin{equation}
\langle\sigma_{t+1}\rangle=P(\sigma_{t+1}= 1)-P(\sigma_{t+1}=-1)=\frac{\gamma}{t}\langle X_t\rangle.
\label{sig}
\end{equation}
Here, 
\begin{equation}
\gamma=(2p-1)(1+\langle f_0(p,t)\rangle).
\label{gamma}
\end{equation}

One can also obtain
\begin{multline}
\langle\sigma^2_{t+1}\rangle = P(\sigma_{t+1}= 1)+ P(\sigma_{t+1}=-1) \\
= \frac{1}{2}-\frac{3}{2}\langle f_0(p,t)\rangle^2+\langle f_0(p,t)\rangle+\frac{(2p-1)^2}{2}\frac{\langle X^2_t\rangle}{t^2}.
\label{p}
\end{multline}
Applying the condition that probability must not be negetive we get from (\ref{sig}) and (\ref{p}) that if $\frac{df_0(p,t)}{dt}\neq 0$ then $f_0(p,t)\sim \frac{1}{3}$ at $t\gg1$. So, for large time either $f_0(p,t)\sim \frac{1}{3}$ or $\frac{df_0(p,t)}{dt}= 0$.\\
Eq-(\ref{sig}) gives rise to the following recursion for the mean displacement 
\begin{equation}
\langle X_{t+1}\rangle=(1+\frac{\gamma}{t})\langle X_{t}\rangle.
\label{1m}
\end{equation}
From the initial condition one can obtain $\langle X_2 \rangle=4s-2$ and $\langle X^2_2 \rangle=8s^2-8s+4$. So, the solution of (\ref{1m}) is obtained by iteration
\begin{equation}
\langle X_{t} \rangle = \langle X_2 \rangle\frac{\Gamma(t+\gamma)}{\Gamma(2+\gamma)\Gamma(t)}\sim  \frac{(4s-2)}{\Gamma(2+\gamma)}t^{\gamma} \quad \textrm{for} \quad  t\gg 1 .
\label{2m}
\end{equation}
For $s=1$ and $p=1$ one must get the trivial ballistic walk i.e. $\langle X_{t} \rangle = t$. So, the maximum value of $\gamma$ is $1$.\\
For the second moment of the displacement one can obtain the following recursion
\begin{equation}
\langle X^{2}_{t+1} \rangle=\left(1+ \frac{2 \gamma}{t}+\frac{(2p-1)^2}{2t^2}\right)\langle X^{2}_{t} \rangle + D(p),
\label{x2}
\end{equation}
where $D(p)=\frac{1}{2}-\frac{3}{2}\langle f_0(p,t)\rangle^2 + \langle f_0(p,t)\rangle$.\\

Solution of the recursion (\ref{x2}) leads to the mean-square displacement for $t\gg 1$
\begin{equation}
\langle X^{2}_{t}\rangle=Gt^{2\gamma}+G\gamma(2\gamma-1)t^{2\gamma-1} \\
-\frac{R}{2\gamma-1}t+\zeta(\frac{1}{t^\beta}),
\label{hurst}
\end{equation}
where $G=\frac{1}{1-\delta_{\gamma,1}\frac{2p-1}{6}}\left(\frac{8s(s-1)+4}{2\gamma(1+2\gamma)}+\frac{2D(p)}{(2\gamma-1)\Gamma(2+2\gamma)}\right)$,\\
 $R=D(p)+\delta_{\gamma, 1}\frac{(2p-1)^2G}{2}$ and $\zeta(\frac{1}{t^\beta})$ refers to the terms of order $\frac{1}{t^{\beta}}$ ( $\beta \geq 1$) which can be neglected for $t\gg 1$.

\begin{figure}    
\includegraphics[scale=0.2]{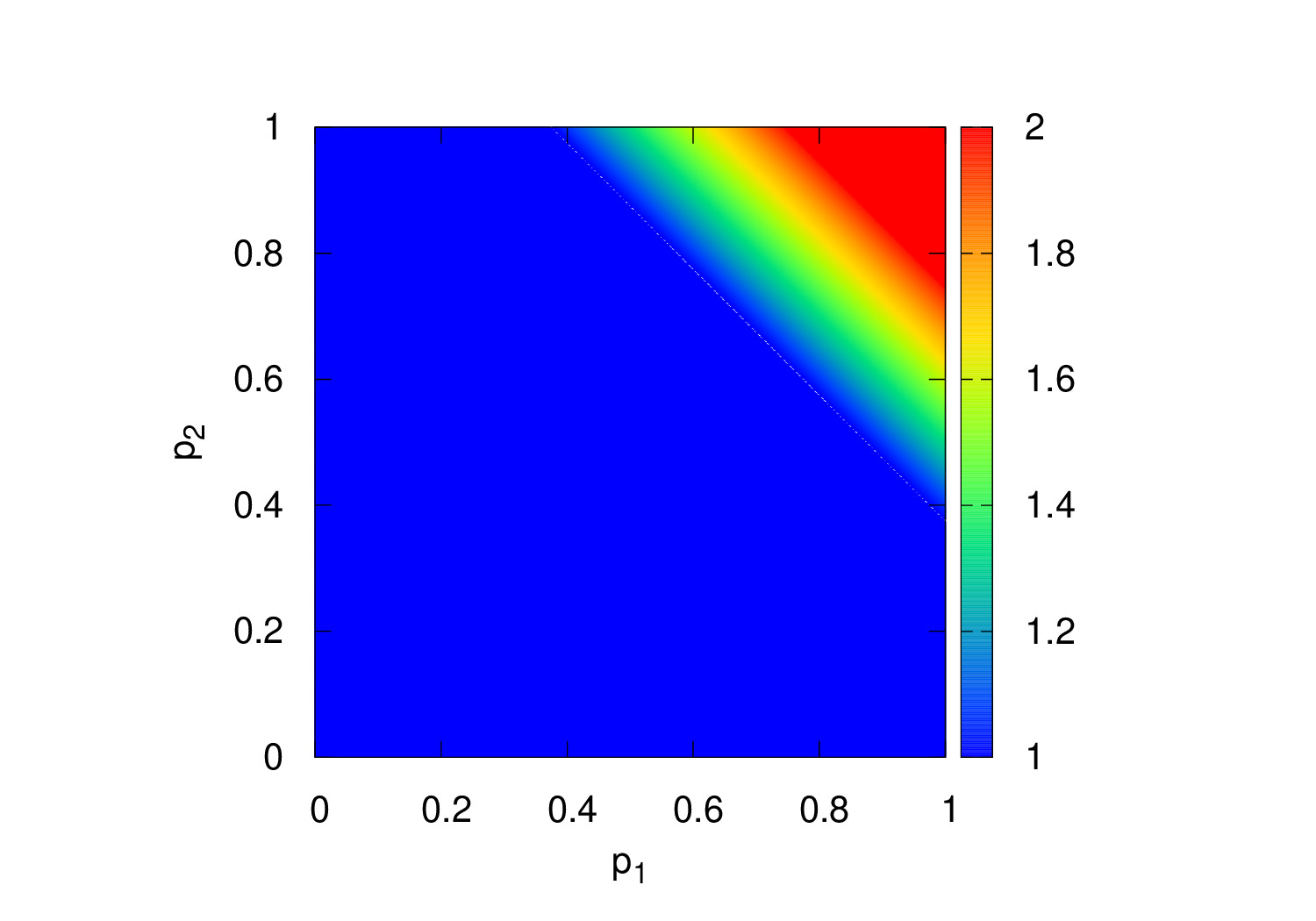}
\caption{Hurst exponent is shown as a function of $p_1$ and $p_2$. Following the line $p_1=p_2$ one can get the phase diagram for single parameter dynamics of RW2MC.}.
\label{H3d}
\end{figure}
Now imposing the condition $\frac{df_0(p,t)}{dt}= 0$ we obtain from (\ref{sig}) and (\ref{p}) that $\gamma$ must be equal to $1$. For $p\geq0.88$ $f_0(p)\sim\frac{1}{3}$ gives $\gamma >1$ , which is not possible. In this case there is no other alternative than $\frac{df_0(p,t)}{dt}= 0$ and $\gamma=1$. So, in (\ref{hurst}) the first term will be the dominant term for $p\geq 0.88$ which gives the hurst exponent $H=2\gamma=2$.  On the other hand, for $p<0.88$ if the value of $\gamma$ is 1, it will violate Eq-(\ref{pc}). In this case one must have $\frac{df_0(p,t)}{dt}\neq 0$ and $f_0(p,t)\sim\frac{1}{3}$ for large $t$ and therefore we obtain from (\ref{gamma}) that $H=\frac{8}{3}(2p-1)$. But for $p\leq0.68$ $2\gamma$ becomes less than one. In this case the third term in (\ref{hurst}) is the dominant term.

So, asymptotically one has
\begin{eqnarray}
\langle X_t^2\rangle &=& G t^2,\quad\quad\quad\quad\quad\quad\,\,\, p\geq 0.88;\\
 \langle X_t^2\rangle &=& Gt^{2\gamma}, \quad\quad\quad 0.68<p<0.88;\\
 \langle X_t^2\rangle &=& \frac{R}{1-2\gamma}t, \quad\quad\quad\quad\quad\! p\leq 0.68.
\end{eqnarray}
for $p<0.68$ (corresponding to $\gamma<0.5$ ) the mean square displacement increases asymptotically linearly in time.\\
for  $0.68<p<0.88$ (corresponding to $\gamma>0.5$) the mean square displacement increases as $\sim t^{2\gamma}$ where $\gamma=\frac{4}{3}(2p-1)$ which is of the same order as the square of the mean but the prefactor is defferent. Therefore the variance becomes superdiffusive. \\
For $p\geq 0.88$ (corresponding to $\gamma=1$) the mean square displacement increases ballistically with a prefactor $G$ where $0<G\leq 1$. The variance still remains superdiffusive but here the internal dynamics is different. Asymptoticaly one has $f_0(p,t)=\frac{2(1-p)}{(2p-1)}$ and $\frac{df_0(p,t)}{dt}=0 $. \\
If we assume two different parameters  $p_1$ and $p_2$ in place of a single parameter $p$  i.e. $\mu^{(1)}_t$  and $\mu^{(2)}_t$ can take positive values ($+1$) with probability $p_1$ and $p_2$ respectively, then following the same procedure it can be obtained that $\gamma=(p_1+p_2-1)(1+\langle f_0(p,t) \rangle)$. Variation of Hurst exponent with respect to $p_1$ and $p_2$ is shown in Fig.\ref{H3d}.    \\


In summary, we have formulated a non-Markovian discrete-time random-walk model RW$n$MC subjected to long-range
time memory correlations. The case for $n=2$ is discussed in detail and the first two moments are exactly calculated. The problem can be studied with 2 parameters $p_1$ and $p_2$ for which the Hurst
exponent is obtained exactly and the  phase diagram in the $p_1$-$p_2$ plane
is fully characterized. The transitions from diffusive to superdiffusive behavior are also described in detail.

In the context of the present work it may be noted that classical random walk with long term memory has been considered before. Such a walk was termed "Elephant random walk"(ERW) which is perhaps the first model to contain memory of  entire history of the walker \cite{erw1}. After that many modifications of the ERW either in the memory pattern or in the decision-making process have been studied \cite{erw2,erw3,cres1,cres2}. Recently quantum walk with long range memory named "Elephant quantum walk" has also been proposed \cite{mo}.

 In ERW walker can choose only one step from its entire history with equal a priori probability i.e. it has only one memory channel. Hence the RW$n$MC may be regarded as a generalisation of the ERW. In RW$n$MC a new paradigm emerges where the memory pattern of different channels can be varied independently. Following this line one can define Alzeimer walk \cite{cres1,cres2,cres3} with multiple memory channels which will obviously further inspire the research on the effect of memory loss. We have constructed a RW2MC in such a way that the dynamics can be expressed by a simple algebric equation (\ref{eq2}). But one can construct a RW2MC in different ways also.  \\
 Connection between the elephant random walk (ERW) and Pólya type Urn model containing balls of two colors and evolving by single drawing has been established earlier \cite{polya1}. Here we have shown that equivalent urn model of the RWnMC presented here is a Pólya type urn containing balls of three colors and evolving by multiple drawing. 
The urn model of RW2MC  presented here has interaction of the type $A+B\rightarrow A+B+C$ i.e. interaction between two drawn balls gives as a product the same two balls plus a new ball. The color of the new ball is determined by a disorder parameter $p$. For $p=1$ (no disorder) one can find some similarity with the urn scheme inspired by the famous rock-paper-scissors game \cite{urnA}.
Interactions of the type $A+B\rightarrow kC$ ($k\in \mathbb{Z}$) or $A+B\rightarrow C+D$ where reactants can annihilate occurs in many population dynamics model with multiple competing species. It is interesting to see whether such model can be described by a RW$2$MC \cite{SS}. In some earlier works \cite{KEM1,KEM2}, walks inspired by dynamical models have been studied leading to some intriguing results. It will be interesting to see if, for example, an opinion dynamics model can be conceived from the present study and compare the results.    
\\ 

\textit{Acknowledgments}.
The author thanks Parongama Sen for useful discussions and crititical reading of the paper. This work is supported by the Council of Scientific and Industrial Research, Government of India through CSIR NET fellowship (CSIR JRF Sanction No. 09/028(1134)/2019-EMR-I)


\end{document}